\documentclass[11pt]{article}

\usepackage{amsmath}
\usepackage{graphicx}
\usepackage{amsfonts}
\usepackage{amssymb}
\usepackage{epsfig}
\usepackage{color}
\usepackage{psfrag}

\newcommand{\EQ}{\begin{equation}}
\newcommand{\EN}{\end{equation}}
\newcommand{\be}{\begin{equation}}
\newcommand{\ee}{\end{equation}}
\newcommand{\bea}{\begin{eqnarray}}
\newcommand{\eea}{\end{eqnarray}}

\begin{document} \setcounter{page}{0}
\topmargin 0pt
\oddsidemargin 5mm
\renewcommand{\thefootnote}{\arabic{footnote}}
\newpage
\setcounter{page}{0}
\topmargin 0pt
\oddsidemargin 5mm
\renewcommand{\thefootnote}{\arabic{footnote}}
\newpage
\begin{titlepage}
\begin{flushright}
SISSA 06/2010/EP \\
\end{flushright}
\vspace{0.5cm}
\begin{center}
{\large {\bf Universal amplitude ratios of two-dimensional percolation from field theory}}\\
\vspace{1.8cm}
{\large Gesualdo Delfino$^{a,b}$, Jacopo Viti$^{a,b}$ and John Cardy$^{c,d}$}\\
\vspace{0.5cm}
{\em ${}^a\,$International School for Advanced Studies (SISSA), via Beirut 2-4, 34151 Trieste, Italy}\\
{\em ${}^b\,$Istituto Nazionale di Fisica Nucleare, sezione di Trieste, Italy}\\
\vspace{1mm}
{\em ${}^c\,$Rudolf Peierls Centre for Theoretical Physics, 1 Keble Road, Oxford OX1 3NP, UK}\\
{\em ${}^d\,$All Souls College, Oxford}\\
\end{center}
\vspace{1.2cm}

\renewcommand{\thefootnote}{\arabic{footnote}}
\setcounter{footnote}{0}

\begin{abstract}
\noindent We complete the determination of the universal amplitude
ratios of two-dimensional percolation within the two-kink
approximation of the form factor approach. For the cluster size
ratio, which has for a long time been elusive both theoretically
and numerically, we obtain the value 160.2, in good agreement with
the lattice estimate $162.5\pm 2$ of Jensen and Ziff.
\end{abstract}
\end{titlepage}

\newpage
{\bf 1.} Universal combinations of critical amplitudes represent
the canonical way of encoding the universal information about the
approach to criticality in statistical mechanics \cite{PHA}. While
critical exponents can be determined working {\em at} criticality,
amplitude ratios characterize the scaling region {\em around} the
critical point. They carry independent information about the
universality class and their determination is in general
theoretically more demanding. Field theory is the natural
framework in which to address the problem, but the usual
perturbative approach is not helpful if one has to work far below
the upper critical dimension $d_c$.

For the best-known example of a geometric phase transition, namely
isotropic percolation ($d_c=6$) \cite{SA}, it was shown in
\cite{DC} how the field theoretical computation of universal
amplitude ratios in two dimensions can be addressed
non-perturbatively exploiting the fact that percolation can be
seen as the $q\to 1$ limit of the $q$-state Potts model \cite{KF},
and that the latter is integrable even away from criticality, in
the scaling limit for $q\leq 4$ \cite{CZ}. Starting from the exact
$S$-matrix \cite{CZ} one can compute the Potts correlation
functions, and from them the amplitude ratios, using the form
factor approach \cite{DC}.

This programme was completed in \cite{DC} for $q=2,3,4$,
recovering the known Ising results and obtaining new predictions
for the three- and four-state Potts model. For percolation,
however, only partial results were obtained, because the
determination of some amplitudes above the percolation threshold
$p_c$ involves the solution of a functional equation which in
\cite{DC} could not be solved for generic values of $q$, in
particular for $q=1$. In this situation, it was observed in
\cite{DC} that a simple parabolic extrapolation to $q=1$ of the
results obtained at $q=2,3,4$ produced for the percolation ratios
results compatible with the available numerical estimates. In
particular, for the ratio of cluster size amplitudes below and
above $p_c$ the extrapolated value 74 essentially coincided with
the central value of the most recent estimate then available
\cite{CJJ}.

On the other hand, the status of the numerical results for the
cluster size ratio (associated to the Potts susceptibility ratio)
was at the time particularly controversial, different authors
having obtained over the years values which spanned two orders of
magnitude \cite{PHA}. Following the appearance of \cite{DC},
Jensen and Ziff communicated a new, very accurate lattice
determination of this ratio, essentially coinciding with the value
$162.5\pm 2$ finally published in \cite{JZ}. A possible
explanation for such a large discrepancy, other than the failure
of the extrapolation (which appeared to work for other ratios),
was discussed in \cite{DBC}, but was ruled out by the full
numerical confirmation the prediction of \cite{DC,DBC} for the
universal ratios at $q=3$ received\footnote{For the case $q=4$, which
is plagued by logarithmic corrections to scaling \cite{CNS,SS}, the issue 
of the precise comparison between field theoretical and lattice results 
for the universal ratios appears still open \cite{CTV,DBC,EG,SBB2}.} 
in \cite{EG,SBB} (see also \cite{CDGJM}).

In this paper we provide the only piece of analytic information
missing in \cite{DC}, namely the solution of the functional
equation at $q=1$, and show that it leads to results for the
percolative universal ratios in complete agreement with the most
recent lattice estimates. In particular, this confirms that the
only problem with the extrapolated value for the cluster size
ratio was the extrapolation itself. When comparing field
theoretical and lattice results (Table~2 below) it must be taken
into account that, with few exceptions, the former are themselves
not exact, since they are obtained truncating the spectral series
for correlation functions to the two-kink contribution. The
remarkable accuracy of this two-particle approximation is however
well known (see e.g. the comparison with the Ising exact results
in \cite{DC}), and is further illustrated by this case.

\vspace{.3cm} {\bf 2.} In \cite{DC} the determination in the
two-kink approximation of the low-temperature spin-spin
correlation function of the scaling $q$-state Potts model ($q\leq
4$) was reduced to that of a function $\Omega_q(\theta)$ entering
the two-kink form factor of the spin field. This function is
characterized by the following properties \cite{DC}:

{i)} is a meromorphic function of $\theta$ whose only singularity
in the strip $\mbox{Im}\,\theta\in(0,2\pi)$ is a simple pole at
$\theta=i\pi$ with residue\EQ
\mbox{Res}_{\theta=i\pi}\Omega_q(\theta)=i\frac{q}{q-1}M,
\label{resOmega} \EN where $M$ denotes the Potts spontaneous
magnetization;

{ii)} is solution of the functional equations \EQ
\Omega_q(\theta)=\Omega_q(-\theta), \EN \EQ
2\cos\frac{\pi\lambda}{3}\sinh\lambda\theta\,\Omega_{q}(\theta)=\sinh\lambda(i\pi+\theta)\,\Omega_{q}(2i\pi-\theta)-\sinh\lambda(i\pi-\theta)\,\Omega_{q}(2i\pi+\theta)\,,
\label{feOmega} \EN with the asymptotic behavior \EQ
\Omega_q(\theta)\sim\exp\left[\left(\frac{2}{3}\lambda-1\right)\theta\right],\hspace{1cm}\theta\to+\infty,
\label{asymp} \EN where $\lambda$ parameterizes $q$ according to
the relation $\sqrt{q}=2\sin(\pi\lambda/3)$.

For $q\leq 3$, where the Potts scattering theory possesses no
bound states, the properties {i)} and {ii)} uniquely identify
$\Omega_q(\theta)$, and then the spin field of the scaling Potts
model\footnote{See \cite{isom,lambda} for the correspondence between
fields and solutions of the form factor equations in integrable
field theory.}.

In \cite{DC} $\Omega_q(\theta)$ was determined only for $q=2,3,4$,
where it takes a simple form. We now show that $\Omega_1(\theta)$
can be obtained taking a mathematical detour in the sine-Gordon
model. For the latter, which is defined by the action \EQ
\mathcal{A}_{SG}=\int\text{d}^2x\text{
}\frac{1}{2}\left(\partial_{\nu}\varphi\right)^2+\mu\cos\beta\varphi,
\EN Lukyanov computed in \cite{Luk} the soliton-antisoliton form
factors \EQ F_{\varepsilon_1\varepsilon_2}^{a}(\theta)=\langle
0|e^{ia\beta\varphi(0)}|A_{\varepsilon_1}(\theta_1)A_{\varepsilon_2}(\theta_2)\rangle,
\qquad\varepsilon_i=\pm 1, \quad\varepsilon_1\varepsilon_2=-1,
\label{ffdef} \EN obtaining a result that in our
notations\footnote{In particular, switching from Lukyanov's
notations to ours involves the replacements $\theta\to-\theta$,
$\xi\to\xi/\pi$, $a\to\beta a$.} reads \EQ
F_{\pm\mp}^{a}(\theta)=-\langle
e^{ia\beta\varphi}\rangle\,\frac{F_0(\theta)}{F_0(i\pi)}A_{\pm}^{a}(\theta),
\label{sgff} \EN \EQ
A_\pm^{a}(\theta)=e^{\mp\frac{\pi}{2\xi}(i\pi-\theta)}[e^{\mp
2i\pi a}I_{a}(-\theta)+I_{a}(\theta-2i\pi)], \label{A} \EN where
$\xi=\pi\beta^2/(8\pi-\beta^2)$, $F_0(\theta)$ is a function on
which we comment below, and $I_{a}(\theta)$ is specified {\em for
real values of $\theta$} and $a\in(-\frac12-\frac\pi\xi,\frac12)$
as \EQ
I_{a}(\theta)=\mathcal{C}\int_{-\infty}^{+\infty}\frac{\text{d}x}{2\pi}\,
W\left(-x-\frac{\theta}{2}+i\pi\right)W\left(-x+\frac{\theta}{2}+i\pi\right)e^{-\left(\frac{\pi}{\xi}+2a\right)x},
\label{Lukint} \EN with \EQ W(\theta)=-\frac{2}{\cosh\theta}\,
\exp\left[-2\int_{0}^{\infty}\frac{\text{d}t}{t}\,\frac{\sinh\frac{t}{2}\left(1-\frac{\xi}{\pi}\right)}{\sinh
t\text{ }\sinh\frac{\xi
t}{2\pi}}\sin^2\frac{t}{2\pi}(i\pi-\theta)\right], \EN and \EQ
\mathcal{C}=\frac14\exp\left[-4\int_{0}^{\infty}\frac{\text{d}t}{t}\,\frac{\sinh^2\frac{t}{4}\text{
    }\sinh\frac{t}{2}\left(1-\frac{\xi}{\pi}\right)}{\sinh
t\text{ }\sinh\frac{\xi t}{2\pi}}\right].
\EN

These results were presented in \cite{Luk} within a framework
known as free field representation, which differs from the usual
approach to form factors based on functional relations. Of course,
this latter approach can be adopted also for the matrix elements
(\ref{sgff}), using as input the sine-Gordon $S$-matrix and the
fact that the soliton is semi-local with respect to
$e^{ia\beta\varphi}$, with a semi-locality phase $e^{2i\pi a}$.
The corresponding functional equations then read \cite{DM} \bea &&
F_{\varepsilon_1\varepsilon_2}^{a}(\theta)=S_T(\theta)F_{\varepsilon_2\varepsilon_1}^{a}(-\theta)+S_R(\theta)F_{\varepsilon_1\varepsilon_2}^{a}(-\theta),
\label{unitarity}\\
&& F_{\varepsilon_1\varepsilon_2}^{a}(\theta+2i\pi)=e^{2i\pi a\varepsilon_2}\,F_{\varepsilon_2\varepsilon_1}^{a}(-\theta),
\label{crossing}
\eea
where
\bea
S_T(\theta)&=&-\frac{\sinh\frac{\pi\theta}{\xi}}{\sinh\frac{\pi}{\xi}(\theta-i\pi)}\,S(\theta),\\
S_R(\theta)&=&-\frac{\sinh\frac{i\pi^2}{\xi}}{\sinh\frac{\pi}{\xi}(\theta-i\pi)}\,S(\theta),
\eea are the transmission and reflection amplitudes; the explicit
form in the present notations of $S(\theta)$ and $F_0(\theta)$ can
be found for example in \cite{DG}, but here we only need to know
that \EQ
F_0(\theta)=S(\theta)F_0(-\theta),\hspace{1cm}F_0(\theta+2i\pi)=F_0(-\theta).
\EN This implies in particular that (\ref{crossing}) is
automatically satisfied by (\ref{sgff}). Since $A_\pm^a$ are
meromorphic functions of $\theta$, also $I_a$, as a linear
combination of $A_+^a$ and $A_-^a$ with entire coefficients, is
meromorphic. It particular, analyticity implies that the property
\EQ I_{a}(\theta)=I_{a}(-\theta), \label{parity} \EN which for
real $\theta$ is apparent in (\ref{Lukint}), extends to the whole
complex plane. Requiring (\ref{unitarity}) leads now to the
equation \EQ 2\cos\left(\frac{\pi^2}{\xi}+2\pi
a\right)\sinh\frac{\pi\theta}{\xi}\,I_{a}(\theta)=\sinh\frac{\pi}{\xi}(i\pi-\eta\theta)\,I_{a}(2i\pi-\theta)-\sinh\frac{\pi}{\xi}(i\pi+\eta\theta)\,I_{a}(2i\pi+\theta)\,,
\label{fe} \EN with $\eta=1$. Making the identifications \EQ
\xi=\frac\pi\lambda,\hspace{2cm}a=-\frac{\lambda}{2}\left(1\pm\frac13\right)+k,\hspace{.6cm}k\in
\mathbb Z \label{id} \EN we rewrite (\ref{fe}) as \EQ
2\cos\frac{\pi\lambda}{3}\sinh\lambda\theta\,I_{a}(\theta)=\sinh\lambda(i\pi-\eta\theta)\,I_{a}(2i\pi-\theta)-\sinh\lambda(i\pi+\eta\theta)\,I_{a}(2i\pi+\theta)\,,
\label{feI} \EN always with $\eta=1$. On the other hand, this
equation coincides with (\ref{feOmega}) when $\eta=-1$. At $q=1$
(i.e. $\lambda=1/2$), however, the sign of $\eta$ becomes
immaterial and (\ref{feI}) exactly coincides with the equation
satisfied by $\Omega_1$.

The functional equation (\ref{feI}) has infinitely many solutions
(a solution multiplied by a $2i\pi$-periodic function of $\theta$
is a new solution) and it remains to be seen whether
(\ref{Lukint}) with the identifications (\ref{id}) and
$\lambda=1/2$ can yield the function $\Omega_1$ relevant for the
percolation problem.

From the known asymptotic behavior (see e.g. \cite{DM}) of the
form factors (\ref{sgff}) one deduces that $I_{a}(\theta)$ behaves
as $\exp\left[\left(a-\frac12\right)\theta\right]$ as
$\theta\to+\infty$, a result which is not obvious from
(\ref{Lukint}) but can be checked numerically. Comparing with
(\ref{asymp}) we see that $I_{a}$ behaves asymptotically as
$\Omega_1$ provided we take $a=-1/6$, corresponding to the lower
sign and $k=0$ in (\ref{id}) with $\lambda=1/2$.

The value $\xi=2\pi$ (i.e. $\lambda=1/2$) falls in the repulsive
regime of the sine-Gordon model in which the only singularity of
the form factors (\ref{sgff}) within the strip
$\mbox{Im}\,\theta\in(0,2\pi)$ is the annihilation pole at
$\theta=i\pi$. Since $F_0(\theta)$ is free of poles in the strip,
the annihilation pole must be carried by $A_{\pm}^{a}$, and then
by $I_{a}$. Any other pole of $I_{a}$ in the strip could not
cancel simultaneously in $A_+^{a}$ and $A_-^{a}$, and then
$\left.I_{a}(\theta)\right|_{\xi=2\pi}$ possesses a single pole at
$\theta=i\pi$ in the strip $\mbox{Im}\,\theta\in(0,2\pi)$, exactly
as it is the case for $\Omega_q(\theta)$ in the Potts model.

Summarizing, the functions
$\left.I_{-1/6}(\theta)\right|_{\xi=2\pi}$ and $\Omega_1(\theta)$
satisfy the same functional relations, have the same asymptotic
behavior and the same singularity structure; then we conclude that
they coincide up to a normalization. Since we know that \cite{DM}
\EQ \mbox{Res}_{\theta=i\pi}F_{+-}^{a}(\theta)=i(1-e^{-2i\pi
a})\langle e^{ia\beta\varphi}\rangle, \EN we read from
(\ref{sgff}), (\ref{A}) and (\ref{parity}) that $I_{a}(\theta)$ has 
residue $i$ on
the pole. Knowing also that the percolative order parameter $P$
(probability that a site belongs to an infinite cluster) is
related to the Potts magnetization as\footnote{The relation
(\ref{P}) is written incorrectly in \cite{DC}, see \cite{DBC}.}
\EQ P=\lim_{q\to 1}\frac{q}{q-1}M, \label{P} \EN and recalling
(\ref{resOmega}), we conclude that $\Omega_1(\theta)$ has residue
$iP$ on the pole, and therefore \EQ
\Omega_1(\theta)=P\left.I_{-1/6}(\theta)\right|_{\xi=2\pi}.
\label{Omega1} \EN The values $\xi=2\pi,\text{ }a=-1/6$ fall in
the range where (\ref{Lukint}) can be used to compute
$\Omega_1(\theta)$ for real values of $\theta$, which is
sufficient for our purposes.

\vspace{.3cm}
{\bf 3.} Near the percolation threshold the relations
\bea
& & S\simeq \Gamma^{\pm}|p-p_c|^{-\gamma},\label{size}\\
& & \xi\simeq f^{\pm}|p-p_c|^{-\nu},\\
& & P\simeq B(p-p_c)^{\beta},\\
& & \frac{\langle N_c\rangle}{N}\simeq
A^{\pm}|p-p_c|^{2-\alpha}.\label{cluster}\eea define the critical
amplitudes for the mean cluster size, the correlation length, the
order parameter and the mean cluster number per site,
respectively; the superscripts $\pm$ refer to $p_c$ being
approached from below or from above\footnote{We keep the notation
of \cite{DC} where $\pm$ referred to the high/low-temperature
Potts phases; we drop instead the tilde used there on percolation
amplitudes.}. Below we consider both the second moment correlation
length \EQ \xi_{2nd}^2=\frac{1}{4}\frac{\int\text{d}^2x\text{
  }|x|^2g_c(x)}{\int\text{d}^2x\text{ }
  g_c(x)},
\EN and the true correlation length $\xi_t$ defined by \EQ
g_c(x)\sim e^{-|x|/\xi_t},\hspace{1cm}|x|\to\infty, \EN where
$g_c(x)$ is the probability that $x$ and the origin belong to the
same finite cluster. It was shown in \cite{DC} that, in terms of
the Potts kink mass $m$, $\xi_t$ is $1/m$ at $p<p_c$ and $1/2m$ at
$p>p_c$, and that \EQ A^{\pm}=-\frac{1}{2\sqrt{3}\,(f_{t}^+)^{2}}.
\label{amplA} \EN Defining the amplitude combinations \EQ
 R_{\xi}^+=\left[\alpha(1-\alpha)(2-\alpha)A^+\right]^{1/2}f^+,\hspace{1cm}U=4\frac{B^2(f_{2nd}^+)^2}{\Gamma^+},
\label{ratiodef} \EN which are universal due to the scaling
relations $2-\alpha=2\nu$ and $2\nu=2\beta+\gamma$, (\ref{amplA})
together with $\alpha=-2/3$ imply in particular \EQ
R_{\xi_t}^+=\left[\frac{40}{27\sqrt{3}}\right]^{1/2}=0.9248.., \EN
a result recovered from a lattice computation in \cite{Seaton}.
The result for $R_{\xi_{2nd}}^+$ in the two-kink approximation was
computed in \cite{DC} and compares quite well with the lattice
estimate obtained from the combination of the data collected in
Table~1.

\begin{table}[htbp]
\begin{center}
\begin{tabular}{|c|c|c|}
\hline
            & \textbf{Triangular} & \textbf{Square}  \\
\hline
$A^+$   & $-4.37^a$             &  -   \\

$\Gamma^+ $ & $0.0720^b$             & $0.102^b$  \\

$B$   & $0.780^b$            &  $0.910^b$   \\

$2f_{2nd}^+$   & $0.520^b$            &  $0.520^b$  \\
\hline
\end{tabular}
\caption{Lattice estimates of critical amplitudes for site
percolation on triangular and square lattice. The superscripts
\textit{a, b} indicate Refs. \cite{DP,AS}, respectively.}
\label{tab1}
\end{center}
\end{table}

The result (\ref{Omega1}) allows us to complete the two-kink
computation of the universal ratios involving the amplitudes
$f_{2nd}^-$, $\Gamma^-$, $B$. All the other necessary information
was already given in \cite{DC} and here we only recall how the
results for percolation follow from those for the Potts model when
$q\to 1$.

Consider as an example the cluster size $S$. This is  the limit of
the Potts susceptibility divided by $q-1$, and the susceptibility
is in turn the integral on the plane of the connected Potts
spin-spin correlator. At $T<T_c$ the leading large distance
contribution to this correlator is produced by a two-kink state
and is multiplied by $q-1$ (the number of two-kink intermediate
states in the low-temperature spectral sum). There are no other
factors of $q-1$ since in this phase the spin two-kink form factor
$F_1^\sigma(\theta)$ is the product of $\Omega_q(\theta)$ times
another function which is also finite at $q=1$ (see \cite{DC}). At
$T>T_c$ the spin-spin correlator coincides by duality with the
low-temperature disorder-disorder correlator. The latter receives
the leading contribution from a single one-kink state weighted by
the squared disorder form factor
$|F_K^\mu|^2=M|F_1^\sigma(\infty)|$ \cite{DC}. As a consequence,
due to (\ref{P}), also the high-temperature Potts correlator
vanishes as $q-1$ in the percolation limit (the two-kink
contribution behaves in the same way). Summarizing, the factors of
$q-1$ can be explicitly isolated and cancel in the computation of
the percolative critical amplitudes for the cluster size. The same
can be shown for the other amplitudes.

\begin{table}[htbp]
\begin{center}
\begin{tabular}{|l|c|c|}
\hline
            & \textbf{Field Theory} & \textbf{Lattice}  \\
\hline
${A^+}/{A^-}$  & 1              &  $1^a$ \\

${f_t^+}/{f_t^-}$   & 2            &  -   \\

${f_{2nd}^+}/{f_t^+}$   & 1.001            &  -   \\

${f_{2nd}^+}/{f_{2nd}^-}$   & 3.73            &  $4.0\pm0.5^c$   \\

${\Gamma^+}/{\Gamma^-}$   & 160.2               & $162.5\pm 2^d$  \\

$U$   & 2.22             &  $2.23\pm0.10^e$  \\

$R_{\xi_{2nd}}^+$   & 0.926              &  $\approx 0.93^{a+b}$ \\
\hline
\end{tabular}
\caption{Universal amplitude ratios in two-dimensional
percolation. The field theory results in the first two lines are
exact, the others are obtained in the two-kink approximation. The
superscripts \textit{a, b, c, d, e} indicate Refs. \cite{DP,
AS,CJJ,JZ,DAS}, respectively.} \label{tab2}
\end{center}
\end{table}

The field theoretical results for the complete list of {\em
independent}\footnote{In \cite{DC} the ratio
$R_c=4(R_{\xi_{2nd}}^+)^2/U$ was considered instead of $U$. The
result $R_c$= 1.56 we obtain should be compared with $\approx
1.53$ following from Table \ref{tab1}.} ratios involving the
amplitudes (\ref{size}--\ref{cluster}) are summarized in Table
\ref{tab2} together with the most accurate lattice estimates. As
remarked above, the comparison confirms in particular the
effectiveness of the two-particle approximated form factor results
in integrable field theory.

\vspace{1cm} \textbf{Acknowledgments.} JC was supported in part by
EPSRC Grant EP/D050952/1. GD and JV were supported in part by ESF Grant INSTANS and by MIUR Grant 2007JHLPEZ.


\begin{thebibliography}{99}
\bibitem{PHA} V. Privman, P.C. Hohenberg and A. Aharony, Universal critical-point amplitude relations, in ``Phase transitions and critical phenomena'', Vol.~14, C. Domb and J.L. Lebowitz eds, Academic Press, New York, 1991.

\bibitem{SA} D. Stauffer and A. Aharony, Introduction to percolation theory (2nd ed.), Taylor \& Francis, London, 1992.

\bibitem{DC} G. Delfino and J. Cardy,  Nucl.Phys. B 519 (1998) 551 [arXiv:hep-th/9712111].

\bibitem{KF} P.W. Kasteleyn and E.M. Fortuin, J. Phys. Soc. Jpn. Suppl. 26 (1969) 11; Physica 57 (1972) 536.

\bibitem{CZ} L. Chim and A.B. Zamolodchikov, Int. J. Mod. Phys. A 7 (1992) 5317.
\bibitem{CJJ} M. Corsten, N. Jan and R. Jerrard, Physica A 156 (1989) 781.

\bibitem{JZ} I. Jensen and R. Ziff, Phys. Rev. E 74 (2006) 020101(R) [arXiv:cond-mat/0607146].

\bibitem{DBC} G. Delfino, G.T. Barkema and J. Cardy, Nucl. Phys. B 565 (2000)
  521 [arXiv:cond-mat/9908453].

\bibitem{EG} I.G. Enting and A.J. Guttmann, Physica A 321 (2003) 90.

\bibitem{SBB} L.N. Shchur, B. Berche and P. Butera, Nucl. Phys. B 620 (2002) 579 [arXiv:cond-mat/0108005]; Phys. Rev. B 77 (2008) 144410 [arXiv:0801.2719 [cond-mat]].

\bibitem{CDGJM} M. Caselle, G. Delfino, P. Grinza, O. Jahn and N. Magnoli, J. Stat. Mech. (2006) 0603:P008 [arXiv:hep-th/05011168].

\bibitem{CNS} J. Cardy, M. Nauenberg and D.J. Scalapino, Phys. Rev. B 22 (1980) 2560.

\bibitem{SS} J. Salas and A.D. Sokal, J. Stat. Phys. 88 (1997) 567 [arXiv:hep-lat/9607030].

\bibitem{CTV} M. Caselle, R. Tateo and S. Vinti, Nucl. Phys. B 562 (1999) 549 [arXiv:cond-mat/9902146].

\bibitem{SBB2} L. Shchur, B. Berche and P. Butera, Nucl. Phys. B 811 (2009) 491 [arXiv:0809.4553 [cond-mat]].

\bibitem{isom} G. Delfino and G. Niccoli, Nucl. Phys. B 799 (2008) 364 [arXiv:0712.2165 [hep-th]].

\bibitem{lambda} G. Delfino, Nucl. Phys. B 807 (2009) 455 [arXiv:0806.1883 [hep-th]].

\bibitem{Luk} S. Lukyanov, Mod. Phys. Lett. A 12 (1997) 2543 [arXiv:hep-th/9703190].

\bibitem{DM} G. Delfino and G. Mussardo, Nucl. Phys. B 516 (1998) 675 [arXiv: hep-th/9709028].

\bibitem{DG} G. Delfino and P. Grinza, Nucl. Phys. B 682 (2004) 521 [arXiv:hep-th/0309129].

\bibitem{Seaton} K. Seaton, J. Phys. A 34 (2001) L759 [arXiv:hep-th/0110282].

\bibitem{DP} C. Domb and C.J. Pearce, J. Phys. A 9 (1976), L137.

\bibitem{AS} A. Aharony and D. Stauffer, J. Phys. A 30 (1997) L301.

\bibitem{DAS} D. Daboul, A. Aharony and D. Stauffer, J. Phys. A 33 (2000) 1113.

\end{thebibliography}
\end{document}